\documentclass{article}

\usepackage{PRIMEarxiv}
\usepackage{url}
\usepackage{hyperref}
\usepackage[utf8]{inputenc}
\usepackage{amsmath}
\usepackage{algorithm}
\usepackage[noend]{algpseudocode}
\usepackage[colorinlistoftodos]{todonotes}
\usepackage{multirow}
\usepackage{enumitem}

\usepackage{xspace}

\begin{document}



\pagestyle{fancy}
\thispagestyle{empty}
\rhead{ \textit{ }} 

\fancyhead[LO]{Running Title for Header}

\date{}

\title{Improving Users’ Passwords with DPAR: a Data-driven Password Recommendation System
}

\author{
Assaf Morag \\
Department of Industrial Engineering \\ 
Tel Aviv University \\
\texttt{assafmorag@post.tau.ac.il}
\And
Liron David \\
Department of Computer Science \\
Weizmann Institute of Science \\ 
\texttt{liron.david@weizmann.ac.il}
\And
Eran Toch\\
Department of Industrial Engineering \\ 
Tel Aviv University \\
\texttt{erant@tauex.tau.ac.il}
\And
Avishai Wool\\
School of Electrical Engineering \\ 
Tel Aviv University \\
\texttt{yash@eng.tau.ac.il}
}

\maketitle


\begin{abstract}
Passwords are the primary authentication method online, but even with password policies and meters, users still find it hard to create strong and memorable passwords. In this paper, we propose DPAR: a Data-driven PAssword Recommendation system based on a dataset of 905 million leaked passwords. DPAR generates password recommendations by analyzing the user’s given password and suggesting specific tweaks that would make it stronger while still keeping it memorable and similar to the original password. We conducted two studies to evaluate our approach: verifying the memorability of generated passwords ($n=317$), and evaluating the strength and recall of DPAR recommendations against password meters ($n=441$). In a randomized experiment, we show that DPAR increased password strength by 34.8 bits on average and did not significantly affect the ability to recall their password. Furthermore, 36.6\% of users accepted DPAR’s recommendations verbatim. We discuss our findings and their implications for enhancing password management with recommendation systems.

\end{abstract}





\section{Introduction}
Passwords are currently the most widely used method for authentication to digital services \cite{Glavin2023}. Despite their many limitations, They are extensively used for protecting sensitive information across numerous services, including web services, online banking, social media, and organizational IT access. Creating robust passwords is essential to the safeguarding of these services. According to industry studies, nearly half of data breaches involved authentication failures \cite{verizon2022,checkpoint2022}. Passwords remain vulnerable because people tend to choose easy-to-remember passwords, which are also commonly used \cite{SHEN2016130}, and tend to reuse the same password in different services \cite{wash2016understanding}. Adversaries then exploit these patterns to craft attacks on systems and individual accounts \cite{suzen2020risk, DUNSAVAGE:2023, CHAUHAN2015203, DOI:2023}. While the use of password managers is increasing, users eschew generated passwords because system-generated passwords are challenging to enter and remember when the manager is unavailable \cite{oesch2022basically,mayer2022users}. Due to their flexibility, simplicity, and wide deployment, passwords are still the preferred method of authentication by users \cite{zimmermann2020password}. Therefore, this study is motivated by the need to help users create stronger passwords, which remains a constant concern for users, organizations, and policy-makers.

Organizations use two primary types of interventions to encourage stronger passwords during password creation: password composition policies and password meters \cite{lee2022password}.
Password composition policies generally rely on blocklists and heuristics and are often found to be less apparent to users \cite{InglesantTrue2010}. Moreover, they do not necessarily produce stronger passwords \cite{komanduri2011passwords, ur2015added,lee2022password} or memorable ones \cite{komanduri2011passwords}. Password meters have more success in encouraging users to create stronger passwords \cite{ur2012does,egelman2013does,ur2017design,golla2018accuracy}. However, the effect of password meters on users' final passwords was shown to be weak \cite{de2014very}. Furthermore, service providers often misuse password meters, implementing them in a way that weakens the final password \cite{lee2022password}. One possible explanation for password-meter limitations is that they fall short in nudging users to change their passwords to stronger ones \cite{renaud2019nudging, kankane2018can}. More interventional approaches, such as providing guidance and detailed explanations to users \cite{FURNELL20181, ur2017design}, had accumulative positive effects on password strengths. These results point to the potential of more concrete and detailed approaches to nudging users in password creation processes, an approach which we take here a step further in generating suggestions to users based on their given password. These processes can be useful not only when asking users to enter new passwords but also in systems that require users to change their password every few weeks or months, a process that users find exceptionally cumbersome \cite{zhang2016revisiting}. 

\begin{figure*}[t]
  \centering
  \includegraphics[width=\textwidth, height=2in]{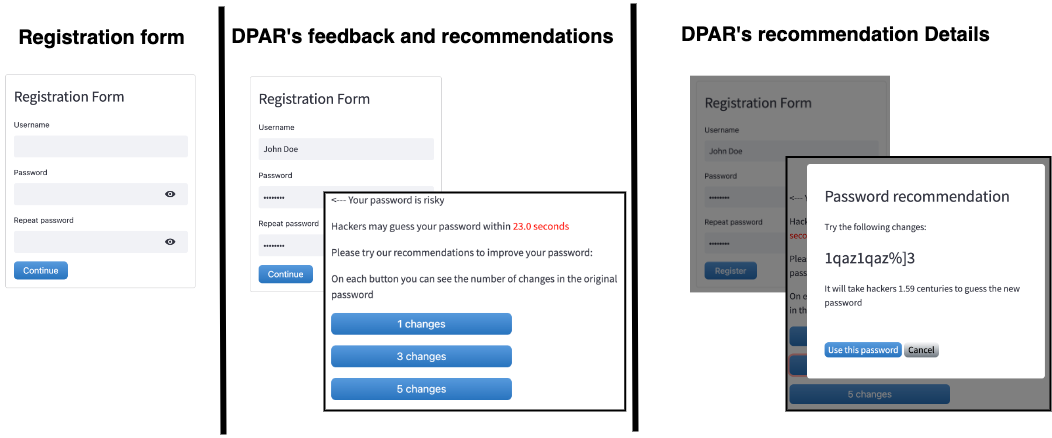}
  \caption{A step-by-step flow of the DPAR user experience, assuming the user initially chose the password ``1qaz1qaz''}
  \label{fig:flow_chart_new_1}
\end{figure*}

In this paper, we introduce \textbf{DPAR -- a Data-driven PAssword Recommendation system}. DPAR takes as input the user's given password and aims to improve it by suggesting specific tweaks to the password that make it more robust but still memorable to the user\footnote{Our source code is publicly available at \url{https://github.com/iWitLab/DPAR/tree/main}.}. To make the recommendations easy to recall, DPAR's recommendations are based on the user's actual given password, in contrast to suggesting completely new random passwords \cite{mukherjee2023mascara}. To do so, DPAR uses a dataset of 905 million leaked passwords \cite{david2022PESrank}, and suggests modified passwords that maximize the similarity to the given password while considerably improving the password strength. The DPAR user experience is embedded in the password creation process, providing the user with a small set of optimized suggestions and a traditional password meter that reflects the password strength to the user (see Figure \ref{fig:flow_chart_new_1}).

We have conducted two studies to evaluate the DPAR recommendation system: (1) A study ($n=317$) designed to measure the correlation between password similarity and perceived usability. our objective was to ascertain whether participants also perceive that a password with fewer modifications is easier to remember. (2) Our main online study ($n=441$) assessed how our recommendations impacted the security and recall of passwords created by participants, and compared it against the state-of-the-art password meter approach. We show that DPAR's recommendations led users to create more secure passwords than just providing feedback on their initial password choice. However, this extra strength did not significantly impact recall metrics: Participants could recall the improved password and reported that the recommendation system was easy to use and helped them create stronger passwords. The key contributions of this work are:
\begin{itemize}[noitemsep,nosep]
\item The design of a password recommendation algorithm that uses a dual analysis of password strength and similarity. This algorithm is intended to synthesize assessments of both security and memorability to proffer optimized password proposals.
\item We introduce and evaluate a configurable user interaction for multiple password recommendations based on the user's given password and show it can significantly improve the strength of users' passwords with comparable recall.
\item We show that users perceive password suggestions most similar to their original password as more usable and thus consistently choose it over other alternatives.
\item We propose an algorithm for generating password recommendations based on a given password, which maximizes both the similarity of the password and its strength. 

\end{itemize}
\section{Background}

\subsection{Password policies}
Typical password policies are designed to make users' passwords more difficult to predict by imposing strict guidelines on the password content. Policies appear in many incarnations: some sites require digits, others require the presence of at least three character classes, some banish certain symbols, and most set varying length minimums and maximums~\cite{bonneau2010password,wang2015emperor,florencio2007large}. A common approach is the LUDS approach, which stands for \emph{Lowercase Uppercase Digits Special Characters}, detailing how many of these classes should appear in the password \cite{wheeler2012zxcvbn}. 

Password policies often pose cognitive burdens on users, which make it harder to create and memorize a password \cite{InglesantTrue2010}. The tension between password security and usability creates competing challenges that influence how passwords are created and used \cite{wash2021prioritizing}. Therefore, users tend to employ simple heuristics to cope with the demands of the policy, such as adding special characters at the end of the password \cite{ur2015added}. These heuristics might paradoxically result in weaker passwords because of their regularity \cite{komanduri2011passwords,guven2022novel}, reducing the barrier for cyber attacks. 

\subsection{Nudging users}
Nudging approaches do not explicitly impose strict policies on users but rather reflect the concept of nudging in behavioral economics to help users make better decisions when creating passwords. Renaud and Zimmerman~\cite{renaud2019nudging} found that a simple textual feedback nudge did not significantly impact password strength. However, adding an incentive in the form of a more extended expiration date for a stronger password resulted in more robust passwords. Kankane et al.~\cite{kankane2018can} found that the warning messages, which point to the ease with which an adversary can compromise the password, were the most effective ways to encourage the creation of stronger passwords. 

Not all measures will work equally well on all users, and giving users several password creation methods might result in more robust and more memorable passwords \cite{yildirim2019encouraging}. Detecting behavioral patterns, such as password reuse \cite{abdrabou2022your}, might require special equipment and infrastructure. Moreover, not every nudging method may result in stronger passwords. When users were asked to change computer-generated passwords, modifying more characters improved memorability but reduced the strength of passwords, as modifications made them easier to crack \cite{huh2015surpass}.

Password meters are a common form of nudging in which the users are presented with information about the password's strength \cite{zimmermann2023hybrid}. Password meters often extend the LUDS approach by including dictionaries, considering l33t speak transformations, keyboard walks, and other heuristics \cite{wheeler2012zxcvbn,guo2018lpse}. Ur et al. developed a data-driven password meter that measures strength and detailed feedback ~\cite{ur2017design}. The meter uses neural network models to calculate the strength of a password and 21 heuristics to generate feedback. While the strength estimation relies on neural network models, which have shown strong estimation capabilities, the feedback and suggestions are ultimately generated by heuristics, which suffer from the same drawbacks as password composition policies.

\subsection{Measuring passwords}

Data-driven approaches to measure password strength aim to simulate how easy it is to crack it, based on heuristics \cite{goodin2013anatomy,hashcat,jtr} or Markov Models \cite{castelluccia2012adaptive,durmuth2015omen,weir2009password}. Neural networks provide a robust way to measure password strength, taking into account the large datasets of leaked passwords that adversaries can easily access \cite{melicher2016fast,david2022PESrank}. Neural networks are trained to generate the next character of a password given the preceding characters of a password. The Password Guessability Service (PGS) \cite{ur2015measuring} provides easy access to the Neural method and some of its predecessors. 

The PESrank model~\cite{david2022PESrank} is based on the observation that when people create a password, they often modify a base word by applying various transformations such as adding a prefix and/or a suffix of digits and symbols before or after the base word, capitalizing some letters of the base word, or substituting letters with visually similar digits or symbols using ``l33t'' translations. PESrank models the probability distributions of these classes (dimensions) of transformations based on a massive dataset of 905 million leaked passwords. 
 
While analyzing the strength of a password is a relatively straightforward task, given a threat model, understanding what makes a password usable requires understanding how users use passwords in the real world. Passwords are not created in isolation, and users take into account their ability to remember and manage multiple passwords for multiple services \cite{woods2017frequently}. Password managers somewhat reduce the cognitive burden of remembering hundreds of passwords, but they do not remove the burden altogether: users still want a memorable password that will work even if they do not have access to their password manager \cite{pearman2019people}. 

Memorability is, therefore, considered the most crucial aspect of password usability. How can we measure memorability? That depends on the usage scenario. Mukherjee et al.~\cite{mukherjee2023mascara} developed a model for generating memorable passwords from scratch based on passphrases \cite{wu2022user}. Levenshtein distance, also known as edit distance, was used in several password-related scenarios, including measuring the impact of privacy policies and meters on modified passwords ~\cite{von2013survival}, determining the similarity between a given password and a dictionary word ~\cite{hu2018password}, or measuring the similarity between a correct and incorrect password during a recall exercise ~\cite{leonhard2007comparative}. However, it is yet unclear whether the Levenshtein distance could serve as a viable approach for generating a usable suggestion based on the user's original password.

\subsection{Password creation systems}
In reviewing the existing literature, we identified two common strategies for password creation mechanisms. The first involves system-generated passwords, while the second accepts an initial user-provided password and strengthens it. In Table \ref{tab:compariso} in the Appendix we summarize the similarities and differences between DPAR and the aforementioned approaches.

System-generated passwords suggest completely new passwords to users and are used in password managers and browsers. For example, Yang et al.~\cite{yang2016empirical} conducted an evaluation of various models that generate passwords for users, specifically focusing on those that employ mnemonic strategies. The study illuminates the advantages of mnemonic-based password generation methods while also providing a comparative analysis of these strategies in terms of their effectiveness and the ease with which passwords can be remembered. Huh et al.~\cite{huh2015surpass} developed the ``Surpass'' system, which generates computer-based passwords while permitting users to modify certain characters to enhance memorability. However, their laboratory studies revealed a trade-off: an increase in character modifications improved password recall but simultaneously decreased security levels, as these passwords were more susceptible to cracking. Mukherjee et al.~\cite{mukherjee2023mascara} introduced the ``MASCARA'' model, aimed at enhancing the recall rate of system-generated passphrases. By assessing the character error rate (CER) of an extensive passphrase dataset, the authors identified three statistically significant signals correlated with memorability. This enabled them to compute the recall rate of passphrases and modify them for improved memorability, albeit with a potential reduction in security.

Forget et al.~\cite{forget2008improving} introduced the concept of Persuasive Text Passwords (PTP) as a way to encourage users to create passwords that are both more secure and usable. This system functions by accepting a password input from the user, and then randomly injecting additional characters into the original password to enhance its strength. Users may select the modified password or to further inject random characters into their original choice. The system was evaluated only in a small-scale lab experiment. Ur et al.~\cite{ur2017design} developed a data-driven password meter (hereafter referred to as the ``CUPS password meter'') that uses a neural network model to assess password strength. It not only provides textual feedback but also adjusts the original password by offering suggestions for enhancement. These suggestions and feedback are derived from password policies. This approach bears the closest resemblance to DPAR. 

Breaking up with the interaction model of password meters, DPAR gives users several password suggestions, mimicking the interaction of classic recommendation systems \cite{harper2015movielens}. This seemingly subtle change in the user interface leads to a considerable shift in how the system operates, as it requires implementing a new method to choose the best passwords that will be presented to the user based on their existing ones. This selection method demands we establish and apply specific criteria to optimize how passwords are picked. In turn, it led to research questions about how the requirements are set and required to rigorously assess the effectiveness of password strength and recall, which were not covered in the existing literature.

\section{DPAR Design and Implementation}

\subsection{Password strength modeling}
In the design of DPAR, we have employed the PESrank model \cite{david2022PESrank} to analyze each password component's strength and the complete password's strength based on a-priori analysis of 905 million leaked passwords. 
As part of the password strength estimation process, a given password is divided into five components (dimensions). If the password begins or ends with a sequence of symbols and digits, the beginning sequence is separated from the password and identified as the prefix. Similarly, the ending digit/symbol sequence is identified as the suffix. The remaining middle part is the mutated base word, which allows for two mutations: specific letters can transform symbols or digits that resemble them (l33t transformations), and lowercase letters can be capitalized. The capitalization pattern and combination of l33t transformations are dimensions 3 and 4. PESrank's base word (after converting to lower case and undoing l33t transformations) is the 5\textsuperscript{th} dimension. For instance, the password ``!1P@ssw0rD2\#'' has a prefix (``!1''), a suffix (``2\#''), a base word (``password''), a l33t list [``@'', ``0''], and a capitalization pattern [0, -1].

Empirical data shows that this 5-dimension decomposition is very well aligned with statistics observed in leaked password corpora. According to an analysis of large leaked password corpora~\cite{david2022PESrank}, 50.24\% of leaked passwords end with digits/symbols and 8.95\% start with digits/symbols---i.e., have a non-empty suffix or prefix. Moreover, 7.66\% use capital letters and 9.86\% use l33t transformations. Finally, while the PESrank model has special case handling for all-numeric or all-symbol passwords (like ``123456'', the \#1 most popular password), these cases cannot occur in DPAR: our password policy requires having at least one letter in the original password.

In the PESrank model, each dimension has its a-priori probability distribution. The global password probability is calculated as a product of the per-dimension probabilities, converted into its rank in the list of possible passwords in decreasing order of likelihood~\cite{dw21jcen}. The rank of the password is its strength since it directly correlates to the time it would take an attacker to crack the password by testing passwords one by one according to the order. PESrank returns $\log_2(rank)$ which measures the password strength in \emph{bits}.


Our primary goal was to enhance the strength of any given password by following the intuitive human manipulations: improving one or more of the non-base dimensions, including prefix, suffix, l33t transformations, and capitalization. The PESrank model provides both strength estimation to the entire password and specific strength estimation to each dimension. This can be leveraged to both explain the password's weaknesses and pinpoint modified dimensions of the password to increase its entire strength. 

\subsection{Password Enhancement} 
The DPAR password enhancement process starts with a user-selected password. The user's input is divided into five dimensions, as explained above. We treat each dimension specifically. We create a list of per-dimension random candidate mutations and then assemble all their possible combinations as a list of password candidates. The base word dimension is left unchanged by our recommendation system. 

A password prefix (or suffix) can be any string composed of 10 digits and 32 keyboard symbols. Our function (GeneratePrefixSuffix in Algorithm~\ref{alg:generation} in the Appendix) always returns three choices for a modified prefix/suffix. It distinguishes between an empty prefix/suffix: the user did not use digits or symbols in the password's beginning (or the end), and a non-empty one. If the original prefix (or suffix) is empty GeneratePrefixSuffix produces three random digit/symbol strings of lengths 1,2,3, respectively. If the original prefix/suffix \emph{str} is not empty, we distinguish between the following scenarios: 
\begin{enumerate}[noitemsep,nosep]
    \item If  \emph{str} is one character long, we randomly replace the character as the length-1 string and randomly append 1 or 2 more digits/symbols to create the length-2 and length-3 strings.
    \item If \emph{str} is two characters long, we randomly replace one character or the other character,  and we randomly add one digit or symbol, producing 2 strings of length 2 and one of length~3.
    \item If \emph{str} is three or more characters long , we randomly select and replace a character three times, producing three modifications and a string of of length~3 or more.
\end{enumerate}

Ultimately, we end up with three sets of prefixes/suffixes that were either replaced or created. We repeat this step 4 times and end up with an array of 12 modified prefixes/suffixes. Note that we only insert between 1-3 changes each time. We do this to control the number of tweaks. Lastly, we add the original prefix (or suffix) to the list of recommendations for a total of 13 options for each dimension.

L33t transformations (replacing letters by digits and symbols that are visually similar) were observed by~\cite{wheeler2012zxcvbn} and shown to be commonly used by~\cite{david2022PESrank}. We used the 14 l33t transformations from Table 2 in \cite{david2022PESrank}. The l33t rank improvement function (GenerateL33t in Algorithm~\ref{alg:generation}) receives the base word and returns one random possible l33t transformation, i.e., when one of a list of 11 lowercase letters [a, e, i, o, s, x, o, p, z, t, g] is located in the base word (these are letters commonly replaced via l33t). Note that some letters have more than one l33t substitution (e.g., `a' can be transformed into `@' or `4'), so there are 14 options. The returned l33t transformation is added to the original l33t transformation for a total of up to 2 options. If the base word doesn't contain any of the letters mentioned above the l33t transformation dimension is left empty. Note that this transformation only replaces letters with digits or symbols, and not the reverse.

The capitalization rank improvement function receives the base word and returns one random capitalization out of all the possible lower-to-upper-case transformations. The returned capitalization is added to the original capitalization transformation (which may be empty) for a total of up to 2 options.  

\paragraph{Recommendation Generation Process}
\label{sec:generationProcess}
In Algorithm~\ref{alg:generation}, we describe the steps taken to create recommendations per each of the non-base dimensions. As described above, each dimensional modification function is activated to create a randomized set of non-base word dimensions. Next, a power set function creates all possible combinations, which are all variations of the original password: up to $13*13*2*2=676$ combinations.

This algorithm produces a dataset of 676 recommendations, where each recommendation consists of the user's base password and at least one modification made by DPAR. Some recommendations in the dataset are modified only in one dimension (prefix, suffix, l33t, capitalization), while others are modified in multiple positions in multiple dimensions. For instance, if the base password of the user was ``amsterdam'', a valid DPAR recommendation could be ``am5terDam\&\#'', in which the user's original choice is modified in 4 positions (``**5***D**\&\#'') spanning 3 dimensions (l33t, capitalization, and suffix).

We calculate the Edit (Levenshtein) distance \cite{levenshtein1966binary} to estimate password similarity between the original password and a given recommendation. The returned value is the minimum number of operations needed to transform one string into the other, where an operation is an insertion, deletion, or substitution of a character. The Levenshtein distance can be, at most, the sum of the lengths of the original and suggested password. 

For each recommendation, DPAR calculates the password strength and  Levenshtein distance. We end up with a list of candidate passwords annotated by their password strength (denoted as PS) and Levenshtein Distance (denoted as LD). We exclude from the list of candidates any passwords whose strength is less than that of the original password. 

We sort all passwords based on their Levenshtein Distance, with ties broken arbitrarily. For each LD value, we retain the password with the highest strength in $Cand[LD]$ and discard the rest. DPAR displays a choice of 3 password replacement recommendations (recall Figure~\ref{fig:flow_chart_new_1}), which are selected by pressing a botton. The upper button leads to a random selection between $Cand[1]$ and $Cand[2]$. The middle button leads to a random selection between $Cand[3]$ and $Cand[4]$, and the bottom button leads to a random selection between $Cand[5]$ and $Cand[6]$.


\vspace*{-2ex}
\subsection{User Experience}
As illustrated in Figure \ref{fig:flow_chart_new_1}, our user interface consists of three screens: the initial landing screen, the feedback and recommendation selection screen, and the specific recommendation screen.

The initial landing screen is divided into two sections. Our system instructs users to create a password with at least eight characters, one letter, and one digit, mimicking a basic password complexity policy. After the user has entered the password, DPAR presents textual feedback based on the password strength, and optionally also three buttons representing our recommendations. The feedback and its text color are determined by the password strength (PS) calculated by the PESrank model. Passwords with $0<PS\le 29$ bits receive the textual feedback ``weak'' colored in red, those with $29<PS\le 37$ bits receive the textual feedback ``fair'' colored in yellow, and those $PS>37$ bits receive the textual feedback ``strong'' colored in green. For instance, ``Your password is weak'', when the word weak is colored in red, as depicted in Figure \ref{fig:flow_chart_new_1}. 

We also estimated how long it would take an attacker to guess the password and placed the color-coded estimate in the feedback. The color coding of the time estimates is aligned with the ``weak/fair/strong'' categorization described above; namely, if the password is found to be ``weak'', the user interface displays the text ``Hackers may guess your password within 4 minutes'', where ``4 minutes'' is colored in red.
We estimated the time (denoted as $T$) it will take a hacker to guess the password by
$\textit{T}= \frac{2^{PS}} {CR}$,
where as before PS is the Password Strength measured in units of bits, and CR is the estimated crack rate.
As CR we used the value of 3.6 million guesses per second, which is aligned with various values other studies used, including 1/2 million guesses per second \cite{di2021password}, 3 million guesses per second, but also reports of systems that can guess up to a few billion guesses per second \cite{ur2015measuring}.

We aimed to avoid burdening the users with too many recommendations on the screen, so we chose to display three buttons with improvement recommendations. 
When a recommendation button is pressed, a window appears and presents an explanation of the recommendation, as shown on the right side of Figure \ref{fig:flow_chart_new_1}. The description includes the recommended password and an estimation of how long it will take a hacker to guess it. The recommendation details screen also includes two buttons. One button says `Use our recommendation'---clicking this button injects the recommendation into the form and closes the details window. The second button says `cancel', and it simply closes the recommendation window.

\vspace*{-2ex}
\subsection{DPAR inputs and configuration}
DPAR is designed to generate a strengthening recommendation to any input password and accepts an input that consists of letters,  digits, and symbols of any length. In case the input password consists of letters only, DPAR will generate recommendations for all the dimensions (prefix, suffix, l33t, and capitalization). For the case of an all-digits or all-symbol input password, DPAR's back end will recognize the whole password as a base word and generate prefix and suffix dimensions only. However, this situation is forbidden by the front-end code that requires at least one letter in the password (Recall Figure~\ref{fig:flow_chart_new_1}).
DPAR can be easily configured to impose limitations and requirements on the user input, such as minimum number of characters, usage of at least one letter or one digit or one symbol or a combination of the above.
DPAR optimization algorithm can also be configured to require minimum strength of a password and preference of various dimensions, for instance the prefix dimension strength will be prioritized over the suffix dimension strength, this will lead to more options with strong prefix in the final recommendations array. 





\vspace*{-2ex}
\section{Perceived memorability study} \label{sec:memorability}

Our first study tested the relationship between a tweaked password and its perceived memorability. We hypothesized that the closer a password recommendation is to the original password created by the user, it will be perceived as easier to remember. To measure the similarity, we used Levenshtein distance (or edit distance). We expected users to rank passwords as more memorable when the Levenshtein distance from the original password is low. To control for the password's strength, we also asked participants to rank the password strength. However, we expected to find that users' perceptions of password strength are poor, as appears in the literature \cite{ur2016users}.

\vspace*{-1ex}
\subsection{Methodology}
A total of $n=317$ participants successfully completed the study. We recruited participants from Amazon’s Mechanical Turk crowdsourcing service, often used in usable security studies \cite{redmiles2019well}. To participate, individuals had to be in the U.S. (to control for cultural and language differences), at least 18 years old, have a Hit approval rate (\%) greater than 98 for all participants’ hits, and have approved at least 1,000 hits. The institutional ethics review committee authorized the study. Participants were compensated \$0.55 for completing the task, which is the equivalent of living per hour. 

\begin{figure*}[!t]
    \centering
    \begin{minipage}[t]{0.48\textwidth}
        \centering
    \includegraphics[width=0.76\textwidth]{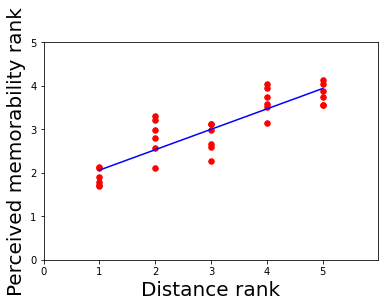}
    \caption{Password Levenshtein distance vs. ranks of perceived memorability, linear regression equation: $y= 0.46X + 1.59$.}
   \label{fig:Lev_Distance}
    \end{minipage}\hfill
    \begin{minipage}[t]{0.48\textwidth}
            \centering
    \includegraphics[width=0.76\textwidth]{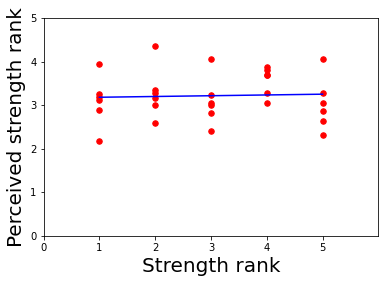}
    \caption{Password strength vs. users' ranks of passwords' strength.}
   \label{fig:Strength_Rank}

    \end{minipage}
\end{figure*}

We presented an online questionnaire to participants. Each participant was presented with three passwords (randomly selected from a pool of 20 passwords). We created five modifications for each password using Algorithm 1, picking one random mutation per distance for distances 1-5. For each variant of the three passwords, we asked participants to rank their strengths and how easy it is to remember. To do so, we asked the participants to order each of the five modified options per password between 1-5: when one is the weakest, five is the strongest, or one is the most difficult to remember while five is the easiest to remember. The questions and answers were presented randomly to avoid biasing the participants' answers with order effects. 

To assess the quality of the participants' data, we performed multiple tests. Initially, 350 participants participated in this study, and 33 (9.4\%) were subsequently excluded from the results due to failures in one or more tests, as follows: 17 participants failed straight-line tests \cite{kim2019straightlining}, 13 submitted incomplete questionnaires, and 3 took an unusually short time to answer the questions. Additionally, we scrutinized the participants' IP addresses to identify and disqualify any addresses listed on blocklists as potential bots, but none of the IP addresses were found on such lists.

We ran a linear regression to analyze how the participants' ranks of password strength and distance correlate with the rank of the password strength based on PESrank and the actual password distance. All statistical tests use a significance level of $\alpha$ = 0.05.

\subsection{Results}

We have measured how users' perceptions of password strength and memorability are related to the actual strength and distance (see Figure \ref{fig:Lev_Distance}). The ranks of passwords' strength were fitted to the PESrank strength according to a generalized linear regression and yielded a poor fit (R\textsuperscript{2}=-0.13). I.e., 
perceptions of password strength were uncorrelated with the actual strength. 

Conversely, the results of ranks of passwords' memorability that were fitted to the Levenshtein distance according to a generalized linear regression yield a good fit (R\textsuperscript{2}=0.68), see linear regression equation in \ref{fig:Lev_Distance}. As is clearly illustrated in the right side of Figure \ref{fig:Lev_Distance}, users feel that passwords with a low distance and few mutations are more likely to be used.

\vspace*{-1ex}
\section{DPAR user study} \label{sec:UserStudy}
Our primary user study evaluated the DPAR system in a scenario resembling a real-world password creation process. We have employed a standard design in password usability studies, similar to several other studies \cite{kelley2012guess, shay2014can,ur2017design}: we created an online study and invited users to register to our online system, guiding them to imagine registering to an important account such as an online banking application or a healthcare account: see the presented instructions in Figure~\ref{fig:imagine} in the appendix.

\vspace*{-1ex}
\subsection{Methodology}
We recruited participants from Amazon’s Mechanical Turk crowdsourcing service in the U.S.. To participate, individuals had to be at least 18 years old, have a Hit approval rate (\%) greater than 98 for all participants’ hits, and have approved at least 1,000 hits. Similar to the memorability study, we paid the equivalent of a living wage to participants, which is \$0.60 for completing Stage~1 and \$0.40 for Stage~2. The Institutional Ethics Committee approved the study. 

The study was divided into two stages: (1) creating a password by registering to our system and (2) measuring password recall by logging in again 48 hours later. In addition to the task at hand, at each stage, we also asked the users to fill out a questionnaire. During Stage 1, participants completed a password creation and recall task and were randomly assigned to a password recommendation feedback variant. Stage 2 occurred at least 48 hours after Stage 1 and included a password recall task and questionnaire. 

We set an upper bound of password strength: Participants whose initial password strength was very strong (exceeded 54 bits) were automatically allocated to the ``feedback only'' group since tweaking such passwords typically yields marginal strength improvements. In the questionnaires, we observed that users with strong initial passwords typically used a password generator. Note that the CUPS Password Meter~\cite{cupsmeter} has similar behavior. To distinguish between them and the original feedback-only group, they were assigned to group number 5. No lower bound of password strength was set.


\paragraph{Stage 1}
Participants were instructed to create a password for a significant account carrying sensitive information. After creating a username and password, they received feedback on the password strength and a recommendation based on their assigned condition (see below). Participants could choose to register without changing their password or to change it. As an immediate ``recall'' step, users logged into the application using their possibly modified password and answered a survey about the password creation process.

\paragraph{Stage 2}
After 48 hours, participants accessed the second stage and logged in to the application. This step was identical to the Stage~1 ``recall'' step: Participants needed to remember their password and username and were then directed to the survey presented in the first stage.

\begin{figure}[t]
 \centering
  \includegraphics[width=0.9\linewidth]{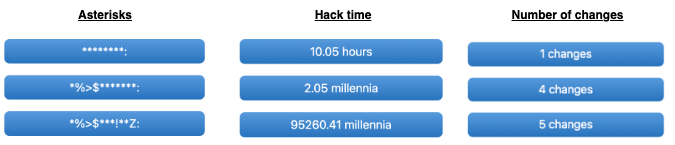}
  \caption{The graphical user interface for the three different feedback+recommendation button variants.}
  \label{fig:Buttons}
\end{figure}

\paragraph{Conditions}
We controlled the study using four independent conditions, as shown in Figure \ref{fig:Buttons}:
\begin{enumerate}[noitemsep,nosep]
    \item Feedback-only: only the password feedback was displayed, with no recommendations. 
    \item Feedback+recommendation (asterisks): the buttons displayed the original password masked with asterisks (*), and only the changes suggested by DPAR were in clear text. 
    \item Feedback+recommendation (number of changes): the buttons displayed the number of changes from the original password.
    \item Feedback + recommendation (hack time): 
 the buttons displayed the new estimated hack time of the recommendation.\footnote{An error in our code caused the ``hack time'' that was displayed for the recommendations to be over-estimated. This visual over-estimation was marginal for weak passwords (strength up to $\approx$40 bits) and more pronounced for stronger password. The final strength values that we use in the analysis are unaffected by the visual error.
}
\end{enumerate}

DPAR output was configured to provide three recommendations. The display screen was set to show each recommendation on a button. We did not clearly display the recommendations to avoid shoulder surfing risk.

Participants assigned to the feedback+recommendations conditions could press any of the three buttons and were shown the recommendation details as in Figure \ref{fig:flow_chart_new_1} (right side). The buttons were ordered according to the number of changes, from the smallest to the largest. They could inject the password into the form by pressing the \textit{`use the recommended password'} button. We randomly assigned the participants to any of the four conditions.

The application was designed to be compatible with both desktop and mobile displays. We chose not to disable any browser features, such as password managers or password storage, to prevent impacting the participants' browsers. This approach also ensured that we did not provide specific guidance that could influence the password creation process.

\paragraph{Survey}
Demographic questions were asked, followed by questions on password creation aids and recall methods. Participants were asked about their experiences during the registration process and to respond on a 5-point scale with questions about the process. They were also asked to share their thoughts and feelings about the process and our recommendations. The questionnaire was adapted to be compatible with both desktop and mobile displays. The questions appear in Appendix 1.

\paragraph{Participants}
The study included 441 participants who completed Stage 1 of the study, and 46.5\% of them finished Stage 2, which is in equivalent proportions to \cite{ur2017design}. Of the participants, 36\% identified as female, 62\% identified as male, and the remaining 2\% identified as another gender or preferred not to answer. The age range of the participants was 19 to 72 years old, with a median age of 36 (mean 37.67, standard deviation 10.82). 

Regarding formal education, 27.9\% of the participants received a high school diploma, 62.4\% graduated with a bachelor’s degree or trade school certificate, 9.7\% graduated with a master’s or Ph.D. degree, and the remaining 0.7\% preferred not to disclose their educational background. Additionally, we requested that participants evaluate their level of self-reported computer proficiency. The results indicate that 34\% of participants ranked their skills as ``very high,'' 41.7\% ranked as ``high,'' 22\% ranked as ``medium,'' 2\% ranked as ``low,'' and 0.3\% ranked as ``very low.'' Notably, there were no significant differences in demographic characteristics between conditions.

We used a random allocation process embedded in the user interface; whenever a new participant accessed the web page, a number between 1 to 4 represented the group affiliation. The distribution of the 441 participants was evenly allocated between the four groups in our study as follows: 24\% were assigned to group 1, 27\% to group 2, and 25\% each to groups 3 and 4. 

To assess the quality of our data, we conducted several tests. Initially, 500 participants participated in this study, and 59 (11.8\%) were subsequently excluded from the results as they all failed to complete the registration task. Specifically, none of them could complete the registration form, which may indicate bot-like behavior. Participants with a very strong initial password, whose strength exceeded 54 bits, were excluded from the statistical analysis ($n=120$, 27.2\%). This was done because there is a less practical need to improve users' passwords with very strong passwords (Mean=64.63, Std=9.07, Median=61.09). The participants were removed with similar proportions from each condition (26 participants from the feedback-only condition and 31 on average from the feedback + recommendation conditions, Chi-Square=3.09, $\textit{p}=0.377$).  

Additionally, the order of the questions and answers in the questionnaire were randomly allocated. We conducted straight-line tests \cite{kim2019straightlining}, reviewed incomplete questionnaires, assessed the time it took to answer the questions, and compared the participants' IP addresses against blocklists to identify potential bots. All the participants, however, passed the aforementioned tests.

\paragraph{Statistical Analysis}
To calculate if there is a statistical difference between the number of participants in each group before and after we extracted the participants with strong initial passwords, we used Chi-Square test for proportions.
To assess statistical significance among the conditions in the study, we analyzed the results using two-way mixed factorial analysis of variance (ANOVA) with the password initial and final password as the within-subjects factor and the two main condition groups, feedback-only and feedback + recommendations as the between-subjects factor. A simple before and after effect test with Bonferroni correction was used to analyze the significance further. All statistical tests use a significance level of $\alpha$ = 0.05.

\begin{figure*}[!ht]
    \centering
    \begin{minipage}[t]{0.48\textwidth}
        \centering
    \includegraphics[width=0.8\textwidth]{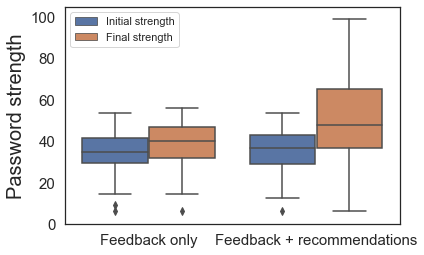}
    \caption{Password strength of feedback-only versus feedback+recommendations participants.}
   \label{fig:box_plot_strength}
    \end{minipage}\hfill
    \begin{minipage}[t]{0.48\textwidth}
            \centering
    \includegraphics[width=0.8\textwidth]{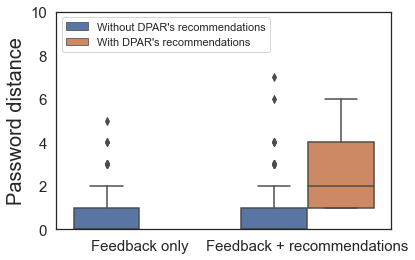}
    \caption{Password similarity between participants who used DPAR's recommendations and those who did not.}
   \label{fig:new_lev_dist}
    \end{minipage}
\end{figure*}

\vspace*{-1ex}
\subsection{Recommendations versus Feedback-Only}
Our first observation from the user study is that among the 238 participants to whom DPAR recommendations were shown, 87 (36.6\%) chose to use the recommendations verbatim (by clicking the `Use our recommendation' button). This is compared to only 7\% of the participants that adopted the suggestions of the CUPS password meter as reported in \cite{ur2017design}. We consider this to be encouraging: it seems that making the recommendations easily accessible through action buttons drives users to strengthen their passwords at much higher adoption rates than when they only receive textual feedback.

\vspace*{-1ex}
\subsection{Password strength and similarity}
\label{sec:StrengthandSimilarity}
We found that choosing a recommended password led users to create a stronger password, while it had no significant impact on the usability measurements in the study. As shown in Figure \ref{fig:box_plot_strength}, the strengths of the initial passwords created by participants from the feedback-only group and the feedback+recommendations group were almost identical. However, the final passwords chosen by the participants who were shown both feedback and DPAR's recommendations were significantly stronger than their initial passwords (from a mean of 36.3 to 51.8, 42.6\%, $p<.001$). In contrast, the final passwords chosen by the participants in the feedback-only condition improved less (from a mean of 34.9 to 38.3, 9.7\%, $p<.001$).




No significant differences were found between different variations of the feedback + recommendation conditions. The final password strength among displaying asterisks ($t=-0.89$, $\textit{p}=0.38$), number of changes ($t=1.57$, $\textit{p}=0.12$), and hack time ($t=0.60$, $\textit{p}=0.55$) all exhibited no significant difference. In other words, it seems that the choice of text displayed on the buttons has little influence on the strength of the final password.

However, a surprising result is that users have a preference for clicking the upper button. Our analysis shows that the upper button (associated with the fewest changes from the original password) was pressed much more frequently (77 times) than the middle (32 times) and bottom (41 times) buttons, even if the types of recommendations were randomized.

As shown in Figure \ref{fig:new_lev_dist}, displaying DPAR recommendation buttons significantly drove users to change their passwords. Participants in the feedback + recommendations condition had a Levenshtein distance average of 2.52, with a minimum of at least one change. On the other hand, participants in the feedback-only condition made only 0.59 changes on average, and participants in the feedback+recommendation condition who did not choose to use the DPAR recommendations exhibited an average distance of 0.52. A t-test for two independent samples between the group that used DPAR's recommendations and the group that did not use them shows a significant effect ($t=-11.095$, $\textit{p}<0.001$).

To test the impact of recommendations on usability, we also measured the time participants took to create their accounts. No notable differences were found among the participant groups' password creation time measurements. The individuals who did not utilize password recommendations had comparable creation time measurements. Those who employed recommendations experienced a minor increase in creation time, but this was not statistically significant. Participants who relied on recommendations and manually typed their passwords into the form exhibited a higher increase in password creation time than the other two groups. However, this increase was still not considered significant.

\vspace*{-1ex}
\subsection{Password Recall}
To test the recall of the passwords, we asked participants to log in immediately after setting the passwords and again in Stage~2, 48 hours after the creation of the passwords.

\begin{table}[t]
\centering
{
\footnotesize
\begin{tabular}{llcc}
   \hline
   Login  & Condition Group& Feedback & Feedback +\\
   Failure&                 & only     & recommendations \\
   \hline
     \multirow{2}{*}{Immediate}& Password kept & 0.0\% & 2.6\% \\
     & Password modified & 1.9\% & 2.1\% \\

     \multirow{2}{*}{48 hours}& Password kept & 22.7\% & 21.6\% \\
     & Password modified& 33.3\% & 19.2\% \\
    \hline
\end{tabular}
}
\caption{Percentage of participants who failed to login after the registration process}
\label{tbl:Table 3}
\end{table}

In the second login attempt, made 48 hours after the password creation, 51.2\% of the feedback + recommendation groups returned to the second part of the experiment, whereas only 41.6\% of the feedback-only group had returned. As appears in Table \ref{tbl:Table 3}, the percentage of failed second login among the participants in the feedback-only group who modified their passwords (33.3\%) was higher than those in the feedback + recommendation groups (19.2\%). While the difference between the proportions was not significant ($\textit{p}=0.357$), they allow us to conclude that the feedback + recommendation group did not underperform in comparison to password feedback methods. 

We control for demographic variables such as age, gender, and education, as well as computer literacy and found no significant difference in login failure rates, see the logistic regression analysis in Table~\ref{tab:logit_results} in the Appendix.

It is also important to stress that the percentage of failed logins on the second attempt, observed after a 48-hour interval, is comparable to the rates of second-attempt login failures reported in other studies: 21.8\%~\cite{ur2017design}, 19\%~\cite{woo2018guidedpass}, 19\%--28\% (across 10 experiment groups)~\cite{tan2020practical}, 20\%~\cite{ye2019empirical} and 28.6\%--79\% (across 7 experiment groups)~\cite{woo2016improving}. 

We observed that participants who created their password using a DPAR recommendation and successfully logged in to the system after 48 hours had an average Levenshtein distance score of 2.86, significantly lower than that of those who failed to log in (Mean=3.17), ($F=4.296$, $\textit{p}<0.001$). This supports our assertion that making fewer changes to the password (i.e., a lower Levenshtein distance) improves password recall beyond the (same) perception reported in Section~\ref{sec:memorability}.

Furthermore, in the questionnaire presented immediately after the password was registered, we inquired about the participants' self-assessed confidence in their ability to recall their passwords, offering the options ``Easy to remember'' and ``Difficult to remember''. We found that among those who reported ``Easy'', only 5\% of those who used one of DPAR's recommendations failed to remember the password 48 after creation, whereas 20\% of those who didn't use DPAR recommendations failed the second login despite claiming it was ``Easy''.
In addition, those who reported ``Difficult to remember'' in fact, found the passwords difficult to remember and had a similarly high fail ratio regardless of whether they used one of DPAR's recommendations (42\%) or not (38\%). Thus, we argue that users' self-assessment of the memorability of DPAR recommendations is aligned with their actual ability to recall the passwords.



\vspace*{-1ex}
\subsection{User interface and attitudes}
\label{sec:buttonChoices}


To better understand the experience the participants had with DPAR, we used an online questionnaire. Table \ref{tab:questionnaire} summarizes the replies to the questionnaire of the participants in the feedback\-+recommendations conditions. More than 80\% of the participants reported they understood the task and found it easy to create the password. On the other hand, only between 28\% and 34\% replied that the experience of working with DPAR was negative (i.e., difficult, confusing, annoying, or redundant). In general, the qualitative feedback and sentiment of the participants are positive and may show users can adopt that password recommendation system.
We also noticed that 36.4\% of participants reported they did not see the recommendations\footnote{The fraction of those who claimed ``I did not see the recommendations'' is perfectly aligned with the number  participants who were allocated to groups 4 and 5 and didn't see any recommendations.}, although they were shown. We mentioned above that 36.6\% of the users used DPAR's recommendations, so those participants saw the recommendations. Therefore, the finding that a significant fraction of participants did not see the recommendations may indicate that there is room to improve the user interface and increase the adoption rate of the recommendations even higher. 

\begin{table}[t]
\centering
{
\footnotesize
\begin{tabular}{lc}
    \hline
    Feedback & Percentage\\ 
    \hline
    I understood the given task & 89.7\% \\
    I felt it was easy to create the password & 80.9\% \\
    The recommendations were helpful & 63.2\% \\
    Creating the password was fun & 46.5\% \\
    I did not see the recommendations & 36.4\%\\
    Creating the password was difficult & 34.6\% \\
    The recommendations were annoying & 30.9\% \\
    The recommendations were redundant & 30.6\% \\
    The recommendations were confusing & 28.3\% \\
    \hline
\end{tabular}
}
\caption{The percentage of participants who agreed or very much agree or agree with the statement}
\label{tab:questionnaire}
\end{table}

For those who reported the password was difficult to remember, we found two main types of explanations: (i) the password was too long, and (ii) the password contained digits or symbols which the users were not accustomed to. Typical feedback was, for instance, \textit{``It was difficult to remember because it added random letters and symbols to my password''}, and \textit{``The password was difficult to remember since I had never used it before and the addition of various numbers and symbols made it hard to remember''}. In most cases, those who reported the password was easy to remember did not add further explanations. But some of the users did recognize that the DPAR only offered minor changes, \textit{``easy because I used a system to make it easy for me to remember and because the suggested change was a small one''}, \textit{``Easy, because it was only changed by a little bit''}, and \textit{``It was easy because it was just a slight adjustment to the password that I had chosen.''}.

\vspace*{-1ex}
\subsection{Validation of the PESrank strength}

\begin{figure}[t]
\centering
  \includegraphics[width=0.83\linewidth]{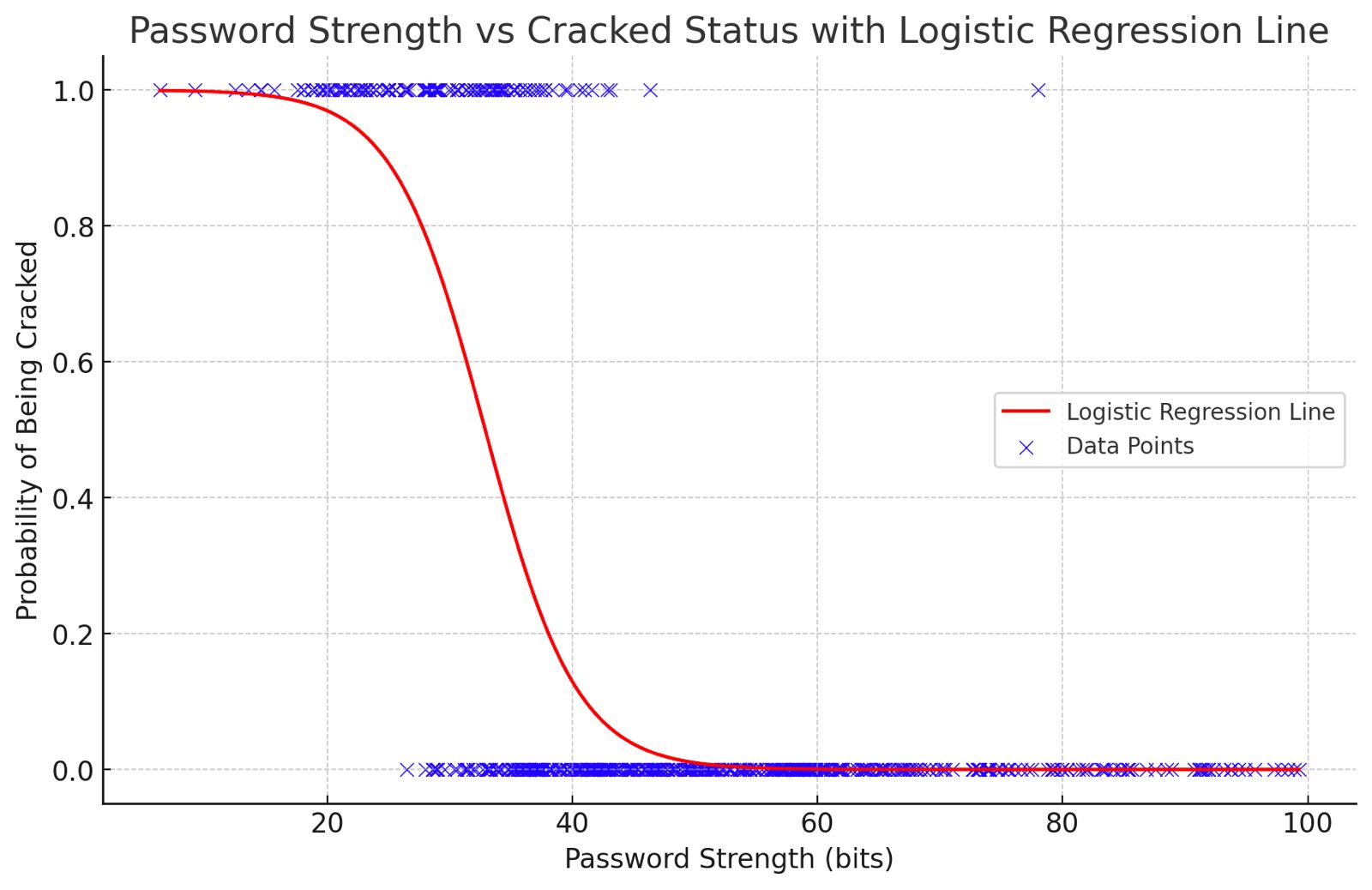}
  \caption{Scatter plot and logistic regression curve of the passwords being cracked by an online hash lookup service as a function of password strength. The regression curve's log-odds of a password being cracked are $\log(p/(1-p))= 8.813 - 0.268X$.  }
  \label{fig:hash_lookup}
\end{figure}

Our data allowed us to validate that the PESrank strength is correlated with stronger passwords. Following common authentication system designs, we did not store the passwords themselves and only kept their SHA256 hash values to let the users log in a second time. Thus, we collected a total of 598 hashes of user-selected passwords. Having the password hashes allows us to validate the correlation between password strengths (as calculated by PESrank) and real password ``crackability''. For this purpose, we attempted to crack the passwords, using 3~popular online hash lookup services \cite{hash_lookup_1, hash_lookup_2, hash_lookup_3} to evaluate if the password was compromised or not. The results are shown in Figure~\ref{fig:hash_lookup}. The graph shows the expected inverse correlation: passwords with higher PESrank scores are less likely to be crackable.

We also performed a logistic regression analysis to examine the influence of password strength, measured in bits, on the likelihood of a password being compromised. 
The model's goodness-of-fit was substantial, with a pseudo $R^2$ value of 0.9064, indicating that PESrank strength can significantly predict the probability of cracking.
The coefficient for password strength was found to be -0.268, underscoring an inverse relationship; as password strength increases, the likelihood of being cracked decreases. 

\vspace*{-1ex}
\section{Discussion}
In this paper, we described the design and evaluation of DPAR, a Data-driven Password Recommendation system that helps users pick up stronger passwords while being able to recall them. We found that users who had used DPAR's password recommendations had ended up with stronger passwords without significantly impacting their ability to recall the passwords. This may suggest that minor modifications to a password that the user initially chose can be recalled similarly to passwords that the user modified. DPAR differs significantly from the approach adopted by password meters \cite{guo2018lpse,ur2017design,forget2008improving}, as it spares users the need to navigate the complexities of password security guidelines or best practices. 

Our user study points to several promising aspects of using recommendation systems in the domain of passwords. DPAR recommendations were chosen by 36.6\% of the participants in the user study, compared to only 7\% of the participants that adopted the suggestions of the CUPS password meter as reported in~\cite{ur2017design}. Most participants who viewed DPAR's recommendations chose the slightest possible change over their original password and the other recommendations. This may indicate that users prefer to make as few changes as possible.

Our findings also point to further complexities that interface designers must face when designing password recommendation systems. For example, users are biased toward clicking on the upper button, even if the order of the buttons is randomized. This phenomenon is a classic example of position effects, where the probability of clinking on a button is influenced by its position \cite{craswell2008experimental}. Each recommendation button represents a trade-off between security and usability, or in behavioral economics terms, a trade-off between cost and benefit; when presented with three recommendation buttons, the upper button is the most immediate to find, and it, therefore, represents higher usability and lower security, the bottom button represents higher security and lower usability, and the one in the middle represents intermediate usability and security. Thus, we speculated that most users would choose the middle ground of usability and security in the cost and benefit \cite{ARIELY1995223}---and this was not the case. We found that most users preferred usability over security. This can emphasize the lack of understanding of security, or perhaps users don't see a lot of benefit in security, thus favoring usability. 

Our findings point to new ways to improve password creation processes. Passwords remain important despite their shortcomings. There is ongoing debate regarding passwords' continued relevance in the face of the growing adoption of alternative authentication solutions \cite{bachmann2014Passwords}, and the declaration of the death of passwords seems premature \cite{quermann2018state,herley2011research,zimmermann2020password}.
However, the usefulness of passwords remains high. Laboratory studies have shown that users highly rate passwords regarding preference, usability, and intention to use, and lowest in terms of expected problems and effort \cite{zimmermann2020password, bovsnjak2019rejecting}. At the same time, while password managers and biometric authentications are becoming more prevalent, they still require passwords as a backup method. Therefore, we argue that the need to help users create more robust and better passwords is a paramount concern that research should address. 

\vspace*{-1ex}
\subsection{Limitations}

When considering the generalizability of our findings, it is essential to understand several limitations of our study. First, we should be aware of some limitations to the usage of m-turk in online studies \cite{tang2022replication}. In the DPAR study, we redirected users to an external application, asking them to complete registration to a website. We also conducted various tests to ensure the quality of our data. Second, the participants in this study were asked to create a password, but unlike a real-life scenario where they would protect a valuable asset, they were not tasked with securing an actual account. This may have led to a reduced motivation to create a secure password compared to those who create passwords to protect important accounts. To mitigate this effect, we instructed the participants to create a password for their high-value online bank accounts or an essential email account, following best practices from password creation studies \cite{kelley2012guess, shay2014can}. In addition, only U.S.-based participants and ASCII-based recommendations were tested in our study, thus the ecological validity is limited to these boundaries, as password composition heavily differs around the world \cite{bonneau2012science, wang2019birthday}.  

Another limitation of our study is the relatively short time in measuring password recall. We measured the recall twice, right after the participants created the passwords and 48 hours later, as was done in a few recent studies \cite{ur2017design, kelley2012guess}, but in real-life scenarios, users must remember passwords for long periods of time: days, weeks, and even months after creating them. Many other factors may affect the recall of these passwords, such as frequency of use and the importance of the account.

To assess the overall contribution of our approach compared to password meters, we note that the mean strength of passwords created without any help was 36.3 bits and improved via DPAR recommendations to a mean of 51.8 bits, while the strength of random and unwieldy passwords generated by password managers was significantly stronger (mean=64.63 bits). Therefore, it is important to note that passwords that are memorable in any way will most probably never reach the strength of fully generated passwords and that the improvements we suggest are inherently limited. 

\vspace*{-1ex}
\subsection{Deploying DPAR and Future Work}

While DPAR helps users improve their passwords and better protect their accounts from hackers, it may also introduce new privacy risks. If not implemented carefully, there's a risk that passwords might be inadvertently exposed. While the risk is not different than the risk imposed by existing password management systems \cite{toch2018privacy}, adding another layer that has access to password content may introduce new risks. These may include the interception of passwords during transmission, unauthorized access due to insufficient security controls, and the possibility of re-identification from anonymized password data. Moreover, the aggregation of password data could inadvertently create a single point of failure, significantly amplifying the impact of any security breach. Given these risks, there is a pressing need for future research focused on privacy-enhancing technologies in the context of password recommendations. 

Existing works on similar problems may point to possible directions. For example, Emmadi et al. implemented a password meter using Fully Homomorphic Encryption (FHE), which performs end-to-end query
privacy for the users' passwords \cite{emmadi2021privacy}. Additionally, developing decentralized approaches that allow password strength evaluation to occur locally on the user's device could eliminate the need to transmit sensitive information, further preserving privacy. The objective of future research would be to provide robust password recommendations while adhering to the standards of user privacy and data protection.

Finally, a potential concern with implementing DPAR as a password recommendation system is that it might prompt attackers to modify their strategies, potentially negating some of the benefits of increased password strength. However, we note that password cracking tools, like hashcat~\cite{hashcat} and John the Ripper~\cite{jtr}, already employ techniques to tweak passwords, and the PESrank measure is designed to simulate such cracking behavior. If DPAR recommendations become popular and crackers adapt, this will change the observed distributions of prefixes, suffixes, l33t, and capitalizations -- and one of PESrank’s advantages is its ability to re-train it with additional corpora. 

\vspace*{-1ex}
\section{Conclusions}
Our study provides a crucial first step in building and implementing password recommendation systems. We have shown that recommendations can add an effective interaction for helping people create stronger passwords while preserving similar recall to password meters. We hope our method can open new avenues for work in making password recommendations useful in real-world systems. Future research should extend the ecological validity of the study, especially when users interact with password recommendations in more diverse contexts. 



\bibliographystyle{plain}
\bibliography{references}

\appendix

\section{Appendix}

\begin{table*}[h]
\centering
{
 \footnotesize
\resizebox{\textwidth}{!}
{\begin{tabular}{l|cccc}
    \hline
    Criteria & DPAR & CUPS \cite{ur2017design} & PTP \cite{forget2008improving} & MASCARA \cite{mukherjee2023mascara}\\ 
    \hline
    Based on user's password & Yes & Yes & Yes & No \\
    
    Password strength model & PESrank & Neural network & Password cracking & Password cracking \\
    Strength is estimated during creation & Yes & Yes & No & No
    \\

    Password usability model & Edit distance, recall tests & Recall tests & Recall tests & Character error rate   
    \\
    Usability is estimated during creation & Yes & No  & No & Yes    
    \\
 Number of recommendations & 3 & 1 & 1 & 3   
    \\
    Recommendation creation mechanism & Data-driven & Heuristics and policies & Random & Random  
    \\
    What in the password is optimized? & Strength and usability & None & None & Strength and usability    
    \\
    Both meter and textual feedback & Yes & Yes & No & No
    \\
    May choose original password & Yes & Yes & No & No
    \\
    \hline
\end{tabular}}
}
\caption{A comparison between DPAR and state-of-the-art password creation systems}
\label{tab:compariso}
\end{table*}

\begin{table*}[t]
\centering
\begin{tabular}{lcccccc}
\hline
\textbf{Variable} & \textbf{Coefficient} & \textbf{Std. Error} & \textbf{z-value} & \textbf{P>|z|} & \textbf{[0.025} & \textbf{0.975]} \\
\hline
const             & 2.6197               & 1.106               & 2.369            & 0.018           & 0.452           & 4.787           \\
Experiment Group  & -0.0114              & 0.191               & -0.060           & 0.952           & -0.386          & 0.363           \\
Age               & -0.0120              & 0.021               & -0.570           & 0.569           & -0.053          & 0.029           \\
Gender            & -0.5259              & 0.417               & -1.261           & 0.207           & -1.343          & 0.291           \\
Computer Literacy & -0.2471              & 0.274               & -0.902           & 0.367           & -0.784          & 0.290           \\
Level of Education & 0.0539              & 0.196               & 0.275            & 0.783           & -0.330          & 0.438           \\
\hline
\end{tabular}
\caption{Logistic Regression Results}
\label{tab:logit_results}
\end{table*}

\subsection{Quastionnaire}
\begin{figure}[h] 
    \centering
    \includegraphics[width = \linewidth]{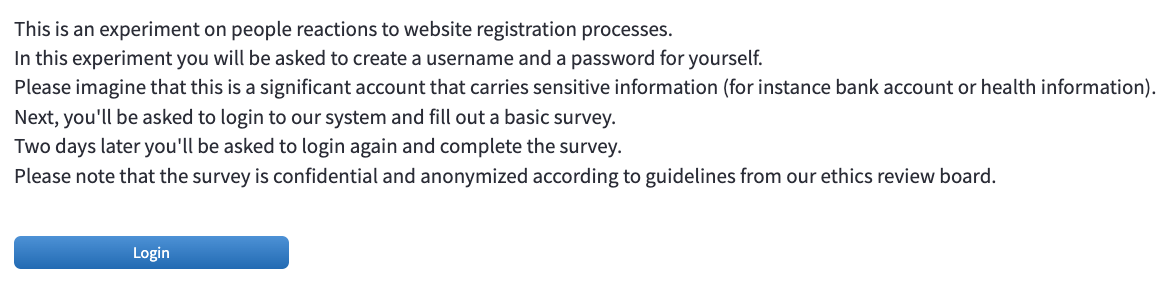}
    \caption{DPAR user study instructions}
   \label{fig:imagine}
\end{figure}

\begin{enumerate}
    \item What is your age?
    \item What is the highest degree or level of school you have completed? If currently enrolled, highest degree received.
\begin{itemize}
\item Some High School
\item High School
\item Bachelor's Degree
\item Master's Degree
\item Ph.D. or higher
\item Trade School
\item I prefer not to answer
\end{itemize}
    \item What gender do you identify as?
\begin{itemize}
\item Female
\item Male
\item Other
\item I prefer not to answer
\end{itemize}
    \item How do you rate your computer literacy and skills?
\begin{itemize}
\item Very high
\item High
\item Medium
\item Low
\item Very low
\end{itemize}
    \item I remembered the username and password I chose:
\begin{itemize}
\item Yes
\item No
\item Only after several attempts
\end{itemize}

    \item I used a password manager, password generator, or browser feature to create my password 
\begin{itemize}
\item Yes
\item No
\end{itemize}

\item I wrote down my password 
\begin{itemize}
\item Yes
\item No
\end{itemize}

\item Please respond per each statement below with the most suitable answer (Strongly agree, agree, natural, disagree, strongly disagree):
\begin{itemize}
\item Creating the password was challenging
\item Creating the password was fun
\item Creating the password was annoying
\item Creating the password was easy
\item Creating the password was difficult
\item Creating the password was confusing
\item The instructions on the registration page were clear
\item The recommendations buttons were visible
\item I understood the task I was given
\end{itemize}

\item Please respond per each statement below with the most suitable answer (Strongly agree, agree, natural, disagree, strongly disagree):
\begin{itemize}
\item The recommendations were helpful
\item The recommendations were redundant
\item I didn't see the recommendations
\item The recommendations were annoying
\item The recommendations were confusing
\end{itemize}

\item The initial password I suggested:
\begin{itemize}
\item I reused a previous password
\item I modified a password I used in the past
\item I created an entirely new password
\item I used a password manager or a password generator to create the password
\end{itemize}

\item The final password that I used:
\begin{itemize}
\item I used one of the passwords that were suggested by the recommendation system
\item I modified my password but didn't use any of the passwords suggested by the recommendation system
\item I didn't use the recommendation system
\item I didn't change the original password I created
\end{itemize}

\item Was it easy or difficult to remember the password, and why? 

\item What did you think about the password recommendations we suggested? 

\item Please write down any comments you might have regarding this study (if you had difficulty understanding the questions, any issues related to the content or the study format, etc.).

\end{enumerate}

\begin{algorithm}[h]
\caption{Recommendations Generation Process}\label{alg:generation}
\begin{algorithmic}[1]
\State \mbox{Input data from the password the user created:}
\State \emph{baseword, prefix, suffix, l33t, capitalization} $\gets$ \emph{password}\;
\State $digSymList \gets [0...9] + [!..?]$\;
\State
\Procedure{GeneratePrefixSuffix}{\emph{str}}     
    \If {$\textit{str is empty}$}
        \State $return \;\; \textit{3 lists of 1, 2 and 3 random digits and symbols} $
    \Else
        \State $return \;\; \textit{3 lists, randomly replace existing digits and symbols} $
    \EndIf
\EndProcedure
\State
\Procedure{GenerateL33t}{$l33t$}
    \State   $l33tcandidate \gets \textit{randomly choose one l33t candidate}$
        \If {$\textit{l33tcandidate is not empty}$}
            \State    $return \; \textit{l33t transformation identifier} $
        \EndIf
\EndProcedure
\State
\Procedure{GenerateCapitalization}{$capitalization$}
    \State  \textit{capitalizationPos} $\gets$
    \textit{random position in the base word}
        \If {\textit{capitalizationPos is not empty}}
            \State    \textit{return \; capitalization position}
        \EndIf
\EndProcedure
\State
\Procedure{powerset}{\textit{prefix', suffix', l33t', capitalization'}}
\State $combinationlist \gets \textit{a powerset of all the groups}$
\EndProcedure

\State \For{$i=1\ldots n_1$}  \;
\State \textit{prefix'} $\gets$ \textit{prefix'} $\bigcup$ GeneratePrefixSuffix(\textit{prefix}) \;
\State \textit{suffix'} $\gets$ \textit{suffix'} $\bigcup$ GeneratePrefixSuffix(\textit{suffix}) \;
\EndFor

\State
\State \textit{prefix'} $\gets$ \textit{original} $\bigcup$ \textit{prefix'}\;
\State \textit{suffix'} $\gets$ \textit{original} $\bigcup$ \textit{suffix'}\;
\State $l33t' \gets \textit{original} \bigcup$ GenerateL33t(\textit{l33t})
\State $capitalization' \gets$
\State \;\;\;\;$\textit{original} \bigcup$ GenerateCapitalization(\textit{capitalization})
\State
\State \textit{DPAR recommendations} $\gets$ powerset(\textit{prefix', suffix', l33t', capitalization'})
\end{algorithmic}
\end{algorithm}

\end{document}